# Neuromatch Academy: Teaching Computational Neuroscience with global accessibility

Tara van Viegen*[1] , Athena Akrami*[2], Kate Bonnen*[3], Eric DeWitt*[4], Alexandre Hyafil*[5], Helena Ledmyr*[6], Grace W. Lindsay*[7], Patrick Mineault*, John D. Murray*[8], Xaq Pitkow*[9,10,11], Aina Puce*[12], Madineh Sedigh-Sarvestani*[13], Carsen Stringer*[14], Titipat Achakulvisut[15], Elnaz Alikarami[16], Melvin Selim Atay[17], Eleanor Batty[18], Jeffrey C. Erlich[19], Byron V. Galbraith[20], Yueqi Guo[21], Ashley L. Juavinett[22], Matthew R. Krause[23], Songting Li[24], Marius Pachitariu[14], Elizabeth Straley[9], Davide Valeriani[25,26,27], Emma Vaughan[28], Maryam Vaziri-Pashkam[29], Michael L. Waskom[3], Gunnar Blohm*+[30], Konrad Kording*+[15], Paul Schrater*+[31], Brad Wyble*+[32], Sean Escola*+[33], Megan A. K. Peters*+[34]

[1] Princeton Neuroscience Institute, Princeton University, Princeton, NJ, USA
[2] Sainsbury Wellcome Centre, University College London, London, UK
[3] Center for Neural Science, New York University, New York City, NY, USA
[4] Champalimaud Research, Champalimaud Foundation, Lisbon, Portugal
[5] Centre de Recerca Matemàtica, Bellaterra, Universitat Autònoma de Barcelona, Barcelona, Spain
[6] International Neuroinformatics Coordinating Facility
[7] Gatsby Computational Neuroscience Unit, Sainsbury Wellcome Centre, University College London, London, UK
[8] Department of Psychiatry, Yale University, New Haven, CT, USA
[9] Baylor College of Medicine, Dept. of Neuroscience, Houston, TX, USA
[10] Baylor College of Medicine, Center for Neuroscience and Artificial Intelligence, Houston, TX, USA
[11] Rice University, Dept of Electrical and Computer Engineering, Houston, TX, USA
[12] Psychological & Brain Sciences, Indiana University, Bloomington, IN, USA
[13] Max Planck Florida Institute for Neuroscience, Jupiter, FL, USA
[14] HHMI Janelia Research Campus, Ashburn, VA, USA
[15] University of Pennsylvania, Depts of Neuroscience and Bioengineering, Philadelphia, PA, USA
[16] Faculty of dentistry, McGill University, Montreal, QB, Canada
[17] Middle East Technical University, Ankara, Turkey
[18] Department of Neurobiology, Harvard University, Cambridge, MA, USA
[19] New York University Shanghai, Shanghai, China
[20] Talla, Inc, Boston, MA, USA
[21] Johns Hopkins University, Department of Biomedical Engineering, Baltimore, MD, USA
[22] Division of Biological Sciences, UC San Diego, La Jolla, CA
[23] Montreal Neurological Institute, McGill University, Montreal, QB, Canada
[24] School of Mathematical Sciences, MOE-LSC and Institute of Natural Sciences, Shanghai Jiao Tong University, Shanghai, P.R. China
[25] Department of Otolaryngology - Head & Neck Surgery, Massachusetts Eye and Ear, Boston, MA, USA
[26] Department of Otolaryngology - Head & Neck Surgery, Harvard Medical School, Boston, MA, USA
[27] Department of Neurology, Massachusetts General Hospital, Boston, MA, USA
[28] IBRO - Simons Computational Neuroscience Imbizo, Cape Town, South-Africa
[29] Laboratory of Brain and Cognition, National Institute of Mental Health, Bethesda, MD, USA
[30] Centre for Neuroscience Studies, Queen's University, Kingston, Ontario, Canada
[31] University of Minnesota, Depts of Psychology and Computer Science & Eng, Minneapolis, MN, USA
[32] Psychology Department, Pennsylvania State University, Philadelphia, PA, USA
[33] Columbia University, Dept of Psychiatry, Center for Theoretical Neuroscience, New York, NY, USA
[34] Department of Cognitive Sciences, University of California Irvine, Irvine, CA, USA
* NMA 2020 Executive Committee, contributed equally

**Abstract**
Neuromatch Academy designed and ran a fully online 3-week Computational Neuroscience summer school for 1757 students with 191 teaching assistants working in virtual inverted (or flipped) classrooms and on small group projects. Fourteen languages, active community management, and low cost allowed for an unprecedented level of inclusivity and universal accessibility.

**Neuromatch Academy**
Traditionally, summer schools have been instrumental in teaching Computational Neuroscience, with training in useful methods and unique networking opportunities. Summer school attendance, to most participants, is a career defining event. However, most neuroscientists never get this opportunity for financial, geographic, or other reasons.

We created Neuromatch Academy (NMA) with the goal of making Computational Neuroscience (and neural data science) summer schools inclusive and globally accessible. Similar to its legacy predecessors, NMA consisted of lectures, tutorials, question and answer sessions, and networking opportunities. Due to the online nature of NMA, we ran pre-recorded lectures, with tutorials in 185 groups or pods and each group led by a teaching assistant. The course was conducted in parallel over 3 major time zones. Students worked on group projects daily to obtain hands-on experience with real data or theoretical projects, providing the opportunity to explore more complex questions or extend the main code-based tutorials. These projects were supervised by 277 volunteer academic mentors. NMA also recreated the career mentorship of traditional summer schools by providing professional development (PD) sessions across multiple time zones on weekends. PD sessions covered practical scientific applications, as well as career panels featuring scientists from both industry and academia. NMA's large size allowed us to introduce students with overlapping interests and the same mother tongue across geographic boundaries, which dramatically increased inclusivity and diversity. Moreover, our ability to attract funds from academia and industry permitted us to offer the school at a low and/or waivable cost to all students.

**Curriculum**
NMA's 3-week curriculum introduced traditional and emerging tools, examining their complementarity, and what they can tell us about the brain. Rather than focusing on tools or algorithms, we aimed to teach approaches to modeling, how to select and use tools, and how to interpret the results [1,2]. The curriculum covered basic to advanced data analysis, modeling and statistical approaches, from biologically realistic to behavioral models, through a carefully-curated progression (Fig. 1A). To ensure coherence within each day's lectures and code-tutorials, each day had a leader who took charge of one of the 15 content days—outlining that day's coverage, prerequisites, and target learning outcomes. To ensure a consistent, comfortable rhythm, all content was delivered in a fixed daily structure that was shifted slightly within each major time zone to accommodate various live events (Fig. 1B; within time zone shifts not shown). After a pre-recorded 'intro' lecture to frame the day's main questions, students explored techniques through hands-on tutorials followed by a pre-recorded 'outro' lecture that showed how the ideas could be extended or applied. The day culminated in a live question-and-answer session with leading experts on the day's topics. Lecturers were selected based on their pedagogical skills, and with consideration of racial and gender diversity, rather than simply for being famous researchers in their field [1].

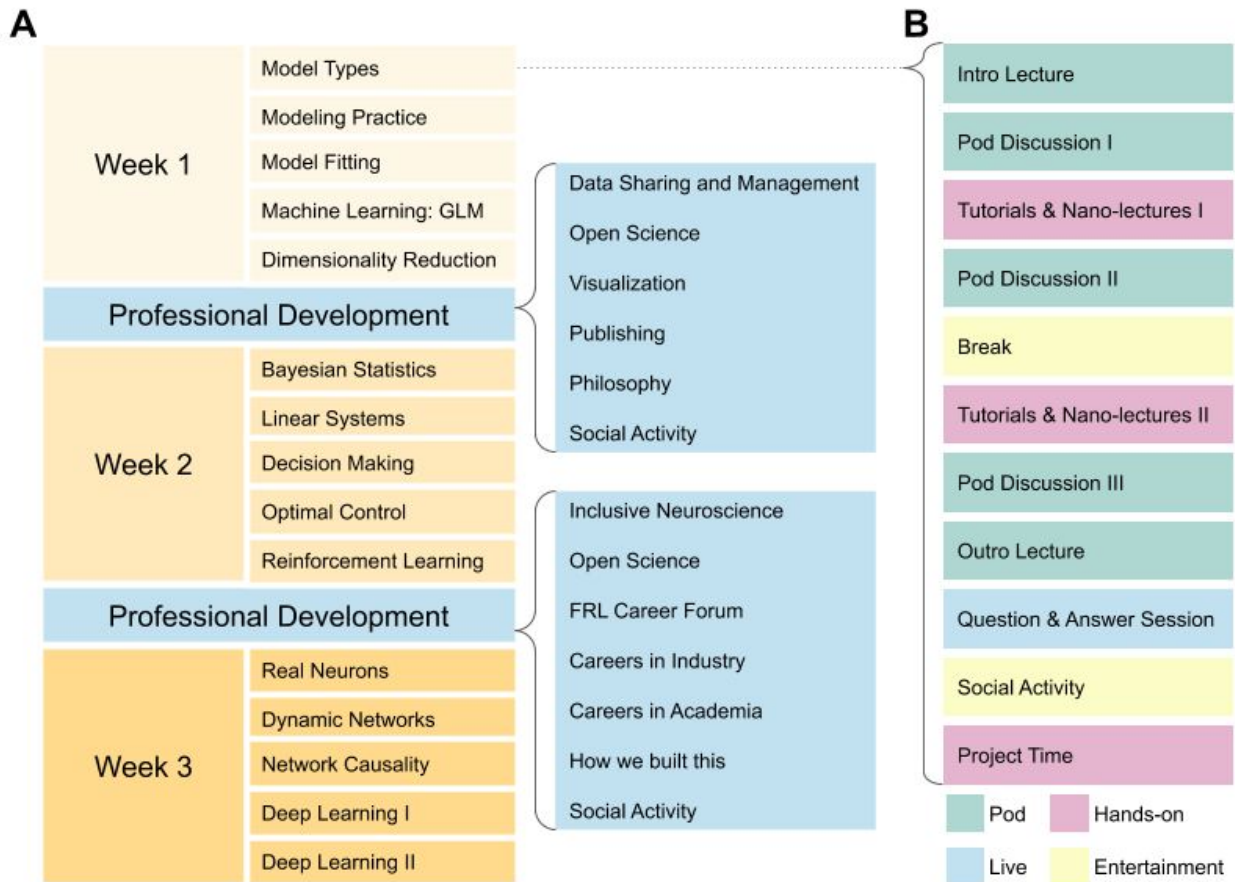

**Figure 1. Curriculum and day-schedule.** The 3-week curriculum of NMA introduced traditional and emerging computational neuroscience tools (A; orange). On the weekends there were several live sessions that focused on professional development (A; blue). Each day consisted of the same schedule as indicated by the dotted line. An example day-schedule is given in B. The ordering of the daily schedule shifted slightly (e.g. shift of social activities or project time) within a time zone to accommodate the live question and answer session in that time zone.

**Tutorial format**

An inverted classroom [3,4] and peer-programming philosophy [5] gave rise to a curriculum consisting of many short hands-on code-based tutorials interspersed with explanatory videos. We used Google Colaboratory[i], a free, browser-based online Python programming environment similar to a Jupyter notebook, which allows code editing and sharing without the need to download or install external software packages. In our tutorials, students never had to produce more than a few lines of code, which gave them more time to engage with other interactive elements—including 'widgets' with sliders or buttons to change the embedded figures in real time—allowing exploration of the applied techniques to build stronger intuitions of models' behavior. Tutorial quality was massively improved by 1) a highly experienced group of code editors, who standardized code across tutorials and 2) a 'dress rehearsal' where the TAs vetted all course materials.

**Group projects**

Students also participated in small group projects (3-5 students) to learn how to apply their new skills. We helped students by providing structured brainstorming, clear expectations and in-depth guidance. Additionally, we offered prepared datasets and suggested projects. Groups were guided by TAs and faculty mentors. After 3 weeks, students gave short talks with slides or videos about their accomplishments. The projects were diverse and some were presented at subsequent Neuromatch Conference 3.0 (neuromatch.io, [6]).

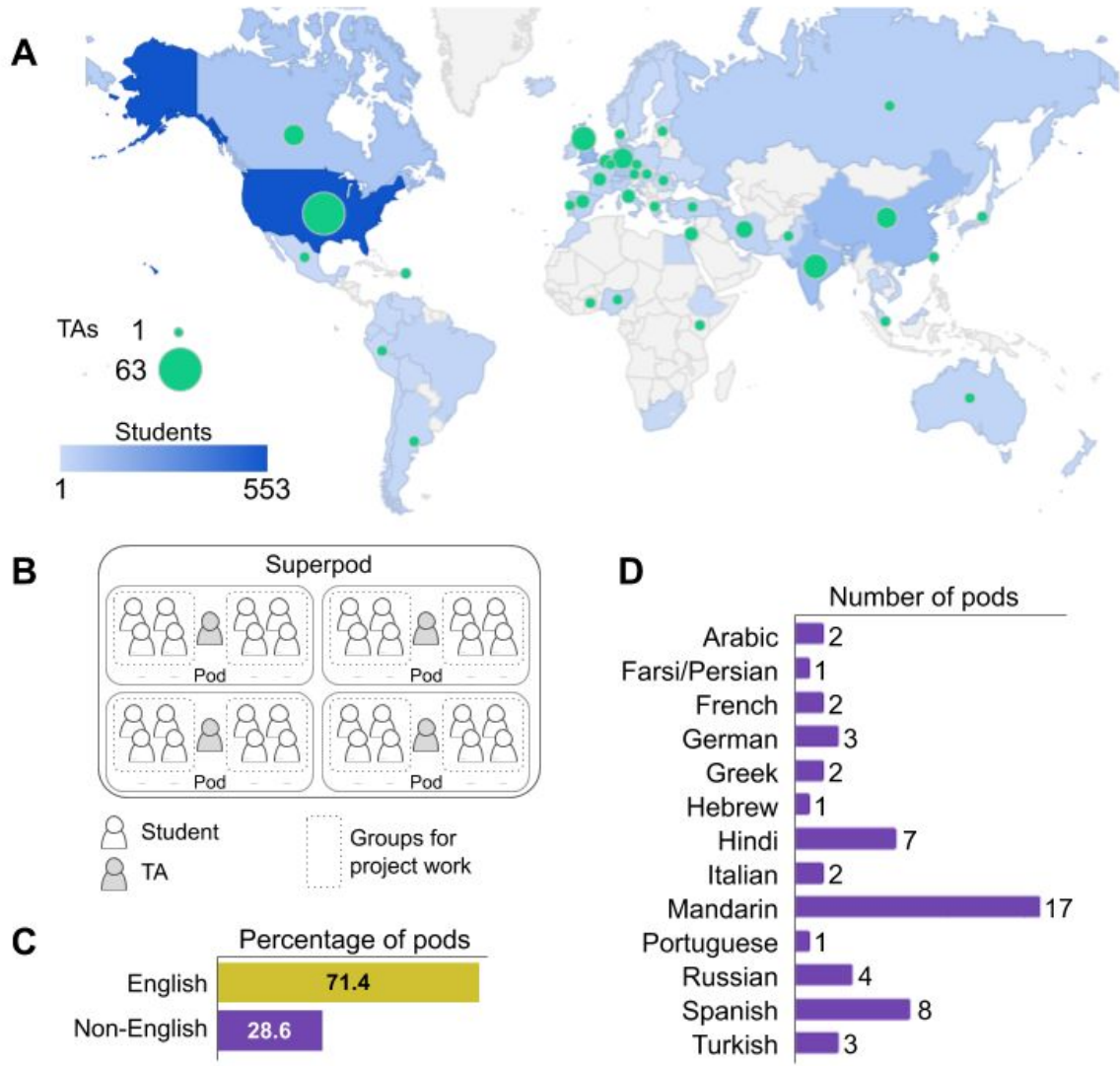

**Figure 2. Student population and school structure.** (A) 1757 students (in blue) and 191 teaching assistants (TAs in green) participated from 64 and 35 different countries, respectively. (B) Pods were created to accommodate small group learning and shifts within and across time zones. Students worked together on project work in groups of 3-5 students (dotted line). Three to four pods formed superpods that shared office hours where students could ask additional questions to TAs and the lead TA. At the end of the three weeks students presented their projects in the superpod. (C) shows the percentage of English and non-English pods and (D) shows the breakdown of the language pods.

**Community building**

We were very active in building community and camaraderie among students. Students were matched into small pods (modal N = 9), each led by a TA based on time-zone preferences, language (Fig. 2C and D), experience and interests (where interests were measured through dataset preferences for group project work; Fig. 2B). The pod structure created intimate groups to facilitate student-centered learning using peer programming [5]. Every pod was given a unique moniker (e.g. 'hairy grasshopper' from the 'coolname' Python package instead of the plain 'pod 123'), for which students created logos (Fig 3A). Other extra-curricular activities covered professional development (Fig. 1A in blue), virtual social hours, yoga classes and karaoke sessions. We also ran algorithmic introductions of students interested in similar topics (neuromatching, [6]) to facilitate networking and enable new collaborations.

About 150 researchers (excluding TAs and project mentors) spanning the ranks of academia from different countries came together as NMA volunteers, driven by the common mission of creating an accessible Computational Neuroscience summer school. Our main tool for communication was Slack[ii] (peaking at 5069 daily messages on July 8th), which has become a mainstream communication tool in science [7]. We used Airtable[iii] to develop a volunteer database of skills and availability, which made task assignment much easier. Credit assignment played a major role in managing this big community effort and we assigned credit through: publishing papers, listing volunteers in all materials and websites, tweeting about some of them, and a 'StarWars' dynamic scroll with contributor names that was screened during live events.

**Assessing NMA's success**

Beyond simply creating a worldwide summer school in three months we surpassed a number of standard metrics. Relative to massive open online courses (MOOCs) NMA had a higher retention rate (86.7% vs 5-10%) [8,9], measured as the percentage of students that had attended >50% of classes. A comment that was repeated often is that students finally felt like they belonged, e.g. 94% of students who responded to the end-of-school survey said they would recommend the experience to a fellow student, and the success of the school was attributed to a wide range of factors (Fig. 3B). Every day, as well as at the end of the school, we polled students and TAs to better understand teaching, learning and community and we had very high rates of survey completion (>90% participation of students and TAs for the final survey). We are thus in a great position to make future summer schools better.

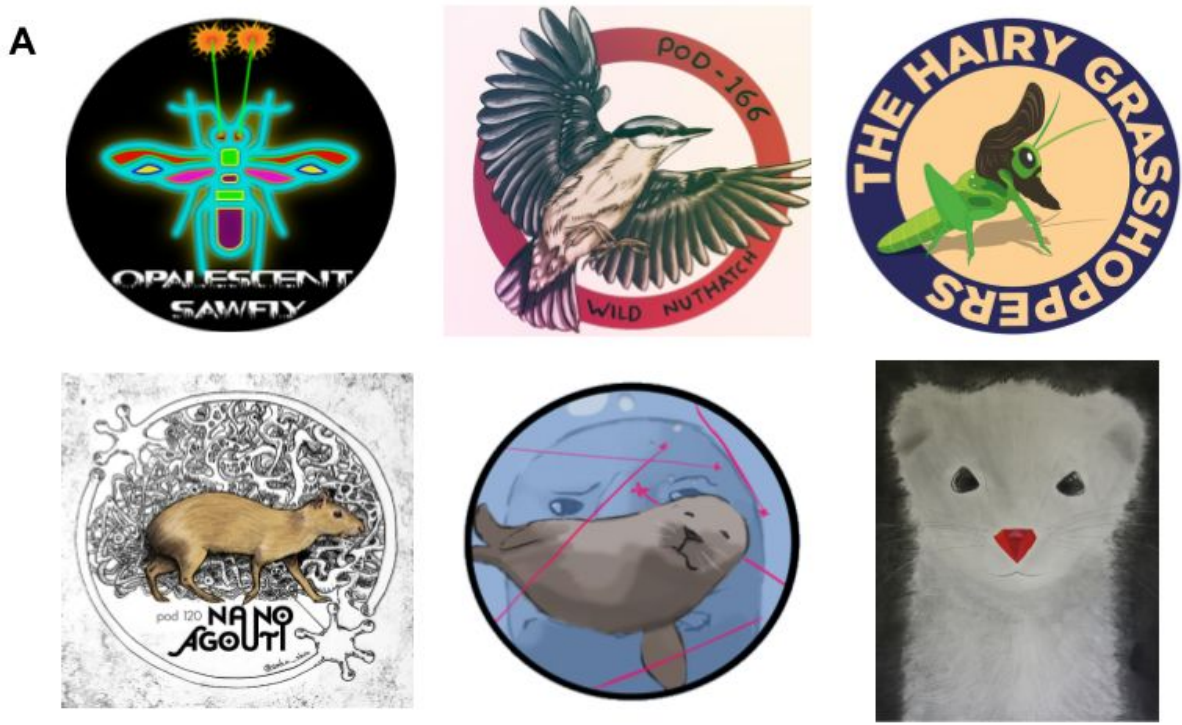
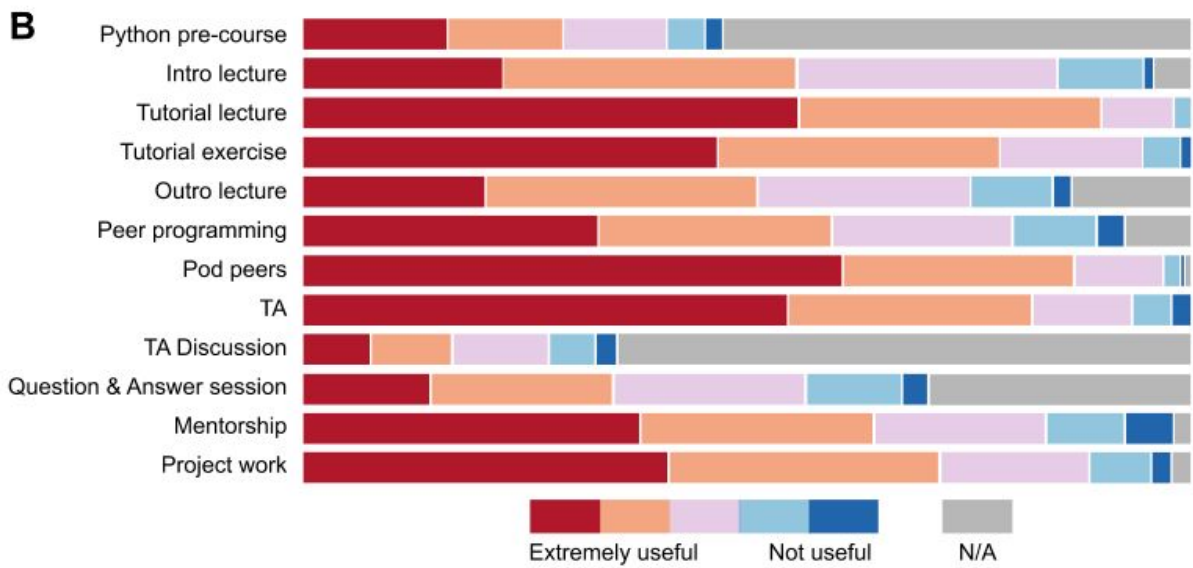

**Figure 3. Students' usefulness scores on different aspects of NMA.** (A) Students and TAs created a pod logo based on their podname. We show a modest selection of the artworks, from the opalescent sawflies, the wild nuthatches, the hairy grasshoppers, the ruby ermines, the nano agouti, the evasive seal and the massive wombat that were shared on social media. Created by: Mohamed Abdelhack, Emily Smith, Kayson Fakhar, Elena Belova, Kyler Mulherin and Maxym Myroshnychenko. (B) Student ratings in end-of-school survey. Extremely useful (leftmost) to not useful is shown from red to blue, with non-applicable (N/A) in gray (rightmost). Pod peers, TAs and tutorial materials scored very high on students' perceived usefulness.

**Building a diverse and inclusive school**

NMA aimed to be diverse and inclusive in many ways. This included: (1) offering language pods so TAs and students could communicate in their mother tongues (despite the NMA language of instruction being English; Fig. 2D); (2) adjusting (already modest) registration fees to reflect national median-wage differences; (3) offering fee waivers; and (4) offering curated closed-captions under pre-recorded materials in English, Spanish, and Mandarin. We created virtual classrooms in which we could communicate, support, and create online communities for our students, TAs, and mentors using the Neurostars forum[iv] from the International Neuroinformatics Coordinating Facility[v] (INCF). We evaluated and maintained the learning atmosphere through Zoom visits by pods from NMA faculty, by monitoring students' daily reflections about their interactions with peers and TAs and by having a Code of Conduct[vi] in place with a rapid-response team to address anonymous reports of violations. We made specific efforts to include students from countries across geopolitical borders, including successfully applying for a last-minute license from the US Office of Foreign Assets Control allowing us to include Iranian resident students and TAs [10]. US sanctions (and those of many other countries) typically prohibit inclusion of Iranian residents (and those from other sanctioned countries) in schools like ours, so we encourage organizers of similar programs to apply for exemptions as early as possible.

All materials were published under a CC-BY license on YouTube[vii] and GitHub[iix]. We used various methods to realize our ideal of open and global access of NMA content, such as mirroring our content to Bilibili for accessibility in China. Our materials are also available as courses through the INCF TrainingSpace. As our materials were all freely and openly available online, students could also participate in our course as so-called Observers (~5000 registered), where students could work through the materials at their own pace. In addition, the coordinated teaching effort of NMA promises to dramatically decrease duplicated effort for educators by optimizing instruction in an inverted classroom format, allowing university faculty to focus more on research while still offering students top-notch education from experts. NMA paved the way for collaborative and coordinated development of teaching materials.

**Conclusion**

To many of us, NMA was an eye-opening experience. By working cooperatively we managed to produce a far better learning experience than we could ever have imagined. We believe that this combination of strong community building, inverted classroom teaching, and a coding-centric curriculum promises major improvements for the future of higher education.

[i] https://colab.research.google.com/
[ii] https://slack.com/
[iii] https://airtable.com/
[iv] https://neurostars.org/
[v] https://training.incf.org/
[vi] http://www.neuromatchacademy.org/code-of-conduct
[vii] https://www.youtube.com/channel/UC4LoD4yNBuLKQwDOV6t-KPw

iix https://github.com/NeuromatchAcademy

**Acknowledgements**

We at Neuromatch Academy are grateful to all of the many, many volunteers who helped make the inaugural Neuromatch Academy in July 2020 a success. A living list of NMA personnel and their roles and contributions can be found at https://www.neuromatchacademy.org/contributors. We are grateful to our evaluation team: Ying-Syuan (Elaine) Huang and Pascal Kropf who kindly provided us with Figure 3b. We are also grateful to our financial sponsors and global partners: Facebook Reality Labs, the Kavli Foundation, the Simons Foundation, the Canadian Institute for Advanced Research, Google DeepMind, the Templeton World Charity Foundation, IEEE Brain, the Wellcome Foundation, the Gatsby Foundation, the Tianqiao & Chrissy Chen Institute, Bernstein Computational Neuroscience Network, the University of Pennsylvania, University of Pennsylvania MindCore, the University of Minnesota, the Columbia University Center for Theoretical Neuroscience, Pennsylvania State University, University of California Irvine, Queen's University, the Howard Hughes Medical Institute Janelia Research Campus, the International Neuroinformatics Coordinating Facility, the Sainsbury Wellcome Center, Rice University, the University of Washington Center for Neuroengineering, Weill Neurohub, Baylor College of Medicine, and the countless small donations of time, money, effort, and love from our friends and colleagues around the world.

**References**

1. Blohm, G., Kording, K. P., & Schrater, P. R. (2020). A How-to-Model Guide for Neuroscience. *eNeuro*, *7*(1). https://doi.org/10.1523/ENEURO.0352-19.2019
2. Blohm, G., Schrater, P., & Körding, K. (2019). Ten Simple Rules for Organizing and Running a Successful Intensive Two-Week Course. *Neural Computation*, *31*(1), 1–7.
3. Lage, M. J., Platt, G. J., & Treglia, M. (2000). Inverting the classroom: A gateway to creating an inclusive learning environment. *The Journal of Economic Education*, *31*(1), 30–43.
4. Baker, J. W. (2000, April). *The "classroom flip": Using web course management tools to become a guide by the side.* Paper presented at the 11th international conference on college teaching and learning, Jacksonville, FL.
5. Williams, L., & Upchurch, R. L. (2001). In support of student pair-programming. *SIGCSE Bull.*, *33*(1), 327–331.
6. Achakulvisut, T., Ruangrong, T., Acuna, D.E., Wyble, B., Goodman, D., & Kording, K. (2020). neuromatch: Algorithms to match scientists. *eLife*.
7. Bottanelli, F., Cadot, B., Campelo, F., Curran, S., Davidson, P. M., Dey, G., Raote, I., Straube, A., & Swaffer, M. P. (2020). Science during lockdown - from virtual seminars to sustainable online communities. *Journal of Cell Science*, *133*(15). https://doi.org/10.1242/jcs.249607
8. Gütl, C., Rizzardini, R. H., Chang, V., & Morales, M. (2014). Attrition in MOOC: Lessons Learned from Drop-Out Students. Learning Technology for Education in Cloud. MOOC and Big Data, 37–48.
9. Yang, D., Sinha, T., Adamson, D., & Penstein, R.C. (2013). "Turn on, Tune in, Drop out": Anticipating student dropouts in Massive Open Online Courses.
10. Ro, C. (2020) How researchers overturned US sanctions on a virtual summer school. *Nature* DOI: 10.1038/d41586-020-02347-9